# The way from high to room temperature superconductivity.

Victor D. Lakhno[*]

Keldysh Institute of Applied Mathematics of RAS, Moscow, Russia
Institute of Mathematical Problems of Biology RAS, Pushchino, Russia

[*]e-mail: lak@impb.ru

**Presently, the main method of producing new high-temperature superconductors is chemical when record temperatures of the superconducting transition $T_c$ are reached due to changes in the chemical structure of a superconductor. This paper deals with a possibility to reach room temperatures of $T_c$ in already available high-temperature superconducting materials by physical methods.**
**It is shown that a superconducting phase corresponds to a Bose condensate moving in a thin subsurface layer and a motionless condensate in volume. We discuss a possibility to set in motion a motionless Bose condensate and enhance the temperature of superconducting transition.**

In paper of 1941[1] L.D. Landau constructed a two-fluid theory of helium II superfluidity as an alternative to the theories by F. London[2] and L. Tisza[3] who associated this phenomenon with Bose-Einstein condensation. These two extreme viewpoints were to a certain extent accommodated by Bogolubov[4] who used a weakly nonideal Bose gas and thus reproduced a phonon-roton spectrum of a two-fluid Landau theory and showed that a superfluid component in this case is a condensate of Bose particles. This, however, did not happen in the case of superconductivity. As was shown by J. Bardeen, L.N. Cooper and J.R. Schrieffer (BCS)[5], Cooper pairs in conventional metals are usually hardly suitable for the role of Bose-particles in view of their huge overlapping. The idea that the superfluidity and superconductivity are relative phenomena was instilled only after the discovery of high temperature superconductivity (HTSC) when in was found that paired states in these materials have small correlation length.

At present there are a lot of candidates for the role of a fundamental Bose particle responsible for HTSC. Competing are however two viewpoints. One of them implies that the pairing mechanism is caused, as in the BCS, by the electron-phonon interaction. According to the other one, pairing of current carriers is caused by magnetic fluctuations[6].

Without going into detail of argumentation in favor of one or the other viewpoint, let us choose the spectrum of a translation–invariant (TI) bipolaron as a spectrum of a fundamental boson responsible for HTSC. The HTSC theory was constructed on the basis of the TI-bipolaron theory[7,8] with the use of the spectrum of its excited states:

$$\varepsilon(\vec{k}) = \omega_0 + k^2/2M , \qquad (1)$$

where $\vec{k}$ is the bipolaron wave vector, $\omega_0$ is the optical phonon frequency, $M = 2m$, $m$ is the effective mass of the band electron which exactly coincides with the roton spectrum in the Landau superfluidity theory[1]: $\varepsilon(\vec{k}) = 0$ corresponds to the ground state of a TI bipolaron.

In this paper we will show that the properties of a Bose gas of TI bipolarons can explain some properties of the superconducting phase in HTSC.

To do this let us consider in greater detail the problem of the motion of Bose condensate of TI bipolarons which determines the superconducting state. In the case of zero temperature T=0 all TI bipolarons are in a condensed state. If such a Bose-condensate moves relative to the crystal lattice of a sample, the total momentum of the Bose-gas relative to the lattice will be equal to $\vec{P}$:

$$\vec{P} = \sum \vec{k}\, m(\vec{k}), \qquad (2)$$

where $m(\vec{k})$ is the Bose function of the distribution of TI bipolarons. In a condensed state each TI bipolaron has one and the same momentum: $\vec{k}_u = M_{bp}\vec{u}$, where $\vec{u}$ is the velocity of a TI-bipolaron in a condensate (i.e. velocity of a Bose condensate), $M_{bp}$ is the bipolaron mass. Accordingly, the distribution function $m(\vec{k})$ in this case will be:

$$m(\vec{k}) = N_0 \Delta(\vec{k} - \vec{k}_u), \qquad (3)$$

where $N_0$ is the number of bipolarons in the condensate which for $T = 0$ is equal to the total number of TI bipolarons $N$, $\Delta(k) = 1$, if $k = 0$ and $\Delta(k) = 0$, if $k \neq 0$. Hence, the total momentum of the Bose-condensate for $T = 0$ will obviously be equal to: $\vec{P} = N_0 M_{bp}\vec{u} = NM_{bp}\vec{u}$.

Now let us consider the case of a nonzero temperature $T < T_c$, where $T_c$ is the temperature of the superconducting transition. In this case some bipolarons are in the excited state. Being in the excited state, a bipolaron can interact with other excitations and defects of the crystal. As a result of such an interaction, the gas of excited states, being in equilibrium with the lattice rests as a whole relative to the lattice. At the same time the excitation gas cannot restrain the condensate part since it cannot exchange momentum with it[9]. As a result, the distribution function of all TI bipolarons in the system of motionless Bose-condensate will have the form:

$$m(k) = N_0 \Delta(\vec{k}) + \left[ \exp\left( \frac{\varepsilon(\vec{k}) + \vec{k}\vec{u}}{T} \right) - 1 \right]^{-1}, \qquad (4)$$

where $\varepsilon(\vec{k})$ is the spectrum of excited states of a TI bipolaron. The formula (4) takes into account that in the reference system in question concerned with the Bose condensate, the excitation gas moves relative to it together with the lattice at a velocity of: $-\vec{u}$.

Substitution of (1), (4) into (2) yields the following value of the total momentum $P'$ of excitations in the system of motionless condensate:

$$\vec{P}' = -M\vec{u}N', \qquad (5)$$

$$N'/V = (MT/2\pi\hbar^2)^{3/2} F_{3/2}(\tilde{\omega}_0/T), \qquad (6)$$

$$F_{3/2}(\alpha) = \frac{2}{\sqrt{\pi}} \int_0^\infty \frac{x^{1/2}}{e^{x+\alpha} - 1} dx, \qquad (7)$$

$$\tilde{\omega}_0 = \omega_0 - Mu^2/2, \qquad (8)$$

where $V$ is the crystal volume, $N' = N - N_0$.

Hence, the total momentum of all the TI bipolarons in the laboratory reference system, i.e. in the system related to the crystal lattice will be equal to:

$$\vec{P} = (N - N')M_{bp}\vec{u} = N_0 M_{bp}\vec{u}. \qquad (9)$$

Expressions (6)-(8) suggest that there is a limiting velocity of the motion of Bose condensate $u < u_c$, $u_c = \sqrt{2\omega/M}$. It follows that the temperature of the superconducting transition $T_c$ (which results from (6) for: $N' = N$) depends on the motion velocity of the Bose condensate and reaches maximum for $\vec{u} = 0$. As the condensate velocity increases, $T_c$ decreases and reaches its minimum: $T_c = 3{,}31 \dfrac{n^{2/3}\hbar^2}{M}$, for $\vec{u} = \vec{u}_c$.

What will happen if the motion velocity of the Bose condensate exceeds its critical velocity, i.e. the inequality $u > u_c$ is satisfied? The integral in (7) in this case does not exist and the steady-state motion appears to be impossible. For $P > P_c = N_0 M_{bp} u_c$ the momentum of excitations starts to be transferred to the condensate restraining it till the condensate velocity becomes equal to $u_c$.

Now let us consider the case when the sample is placed in a magnetic field. In view of Meissner effect, the magnetic field sets in motion the Bose condensate at the subsurface layer of the sample whose thickness is of the order of London penetration depth. Hence, the velocity $\vec{u}$ in a sample in a magnetic field appears to be distributed heterogeneously. Since the magnetic field does not penetrate into the sample, the main mass of the Bose condensate occurring in the sample will be motionless, playing the role of a kind of a 'dark matter'.

It follows that $T_c$ in the subsurface layer differs greatly from that in the bulk of the sample. Thus, for example, for the bipolaron concentration $n_{bp} \approx 2 \cdot 10^{20}$ cm$^{-3}$, $m = m_0$ - electron mass in vacuum; $\omega_0 = 50$ mev, according to (6)-(8), the critical temperature in the bulk of the sample (corresponding to $\omega_0(u=0) = 50$ mev) as compared to its surface value (corresponding to $\omega_0(u=u_c) = 0$) turns out to be nearly three-fold higher. For superconductors with $T_c \approx 100$ K, involvement of the bulk part of the Bose condensate into the motion can three-fold enhance $T_c$, i.e. exceed the room temperature. This phenomenon was probably realized by S.C. Pais[10]. With this aim Pais[10] used an alternating magnetic field which deteriorates the surface superconductor layer and an alternating current which sets the bulk Bose condensate in motion.